\documentclass[aps,prd,nofootinbib,twocolumn,superscriptaddress,floatfix,notitlepage]{revtex4-2}

\usepackage{graphicx}
\usepackage{amsmath,amsfonts,amssymb}
\usepackage{color}
\usepackage{verbatim}
\usepackage{enumitem}
\usepackage{aas_macros}
\usepackage[breaklinks,colorlinks,urlcolor=blue,citecolor=blue,linkcolor=magenta]{hyperref}

\newcommand{\td}{{\rm d}}

\newcommand{\vect}[1]{\boldsymbol{#1}}

\newcommand{\be}{\begin{equation}}
\newcommand{\ee}{\end{equation}}
\newcommand{\bea}{\begin{equation} \begin{aligned}}
\newcommand{\eea}{\end{aligned} \end{equation}}

\def\lsim{\mathrel{\raise.3ex\hbox{$<$\kern-.75em\lower1ex\hbox{$\sim$}}}}
\def\gsim{\mathrel{\raise.3ex\hbox{$>$\kern-.75em\lower1ex\hbox{$\sim$}}}}

\newcommand{\papertitle}{JWST Constraints on Primordial Magnetic Fields}

\begin{document}

\title{\papertitle}

\author{Malcolm Fairbairn}
\email{malcolm.fairbairn@kcl.ac.uk}
\affiliation{Theoretical Particle Physics and Cosmology, King’s College London, Strand, London WC2R 2LS, United Kingdom}

\author{María Olalla Olea-Romacho}
\email{maria\_olalla.olea\_romacho@kcl.ac.uk}
\affiliation{Theoretical Particle Physics and Cosmology, King’s College London, Strand, London WC2R 2LS, United Kingdom}

\author{Juan Urrutia}
\email{juan.urrutia@kbfi.ee}
\affiliation{Laboratory of High Energy and Computational Physics, NICPB, R{\"a}vala 10, Tallinn, 10143, Estonia}
\affiliation{Department of Cybernetics, Tallinn University of Technology, Akadeemia tee 21, 12618 Tallinn, Estonia}

\author{Ville Vaskonen}
\email{ville.vaskonen@kbfi.ee}
\affiliation{Laboratory of High Energy and Computational Physics, NICPB, R{\"a}vala 10, Tallinn, 10143, Estonia}
\affiliation{Dipartimento di Fisica e Astronomia, Universit\`a degli Studi di Padova, Via Marzolo 8, 35131 Padova, Italy}
\affiliation{Istituto Nazionale di Fisica Nucleare, Sezione di Padova, Via Marzolo 8, 35131 Padova, Italy}

\begin{abstract}
Primordial magnetic fields (PMFs) enhance small-scale structure formation through the Lorentz force acting on baryons, boosting the abundance of low-mass halos and their hosted galaxies. We show that the reionisation history calibrated with the UV luminosity function (UVLF) provides stringent bounds: strong PMFs induce a characteristic double reionisation at $z \approx 24$ that is incompatible with CMB measurements of the optical depth, yielding $\sqrt{\left\langle B^2 \right\rangle} < 0.27\,{\rm nG}$ and $< 0.18\,{\rm nG}$ for $n_B = -2$ and $n_B = 2$ respectively at $95\%\,{\rm CL}$ using Planck priors on $\tau$. This establishes early galaxy observables as among the most sensitive probes of PMFs in Gaussian, non-helical scenarios.

\vspace{1em}
KCL-PH-TH/2025-37
\end{abstract}

\maketitle

\noindent\textbf{Introduction  -- } Magnetic fields are widespread across vastly different scales, from planets, stars and galaxies~\cite{2010SSRv..152..651S, Reiners:2012bb, Beck:2008ty, Beck:2013bxa} up to galaxy clusters~\cite{Vogt:2005xf, Osinga:2022tos}. In galaxies, they can be efficiently amplified by astrophysical processes~\cite{Robishaw:2008ti,McBride:2012mm,Bernet:2008qp,Geach:2023enq,Schober:2013aoa,Martin-Alvarez:2021jsh,Hanayama:2005hd,Sur:2012zj}, but all such mechanisms require a pre-existing seed field whose origin remains unknown. This has motivated both astrophysical mechanisms such as the Biermann battery~\cite{1950ZNatA...5...65B} and cosmological scenarios including inflationary magnetogenesis and phase transitions~\cite{Turner:1987bw, Ratra:1991bn,Vachaspati:1991nm,Sigl:1996dm, Enqvist:1993np, Ellis:2019tjf, Balaji:2024rvo, Olea-Romacho:2023rhh, Olea-Romacho:2025qag}. While in galaxies and clusters both remain viable~\cite{Donnert:2008sn, Vazza:2014jga}, the magnetisation of cosmic web filaments~\cite{Govoni_2019} and gamma-ray observations of blazars suggests that fields in intergalactic voids~\cite{Neronov:2010gir} have a primordial origin, with plasma instabilities as an alternative explanation remaining under debate~\cite{Broderick:2011av}.

The impact of primordial magnetic fields (PMFs) on structure formation has been investigated for several decades. Wasserman~\cite{1978ApJ...224..337W} first showed that stochastic magnetic fields could seed galaxy-scale fluctuations, and subsequent work developed a linear perturbation framework for magnetised cosmologies~\cite{Kim:1994zh, Subramanian:1997gi, 2003JApA...24...51G}. Beyond the linear regime, the evolution of magnetised baryons becomes non-linear below the damping scale $\lambda_D$, defined by $v_A/\lambda_D \sim aH$, where $v_A$ is the Alfv\'en speed. Dedicated simulations have recently explored this regime, revealing saturation of small-scale power and modification of the baryonic content of high-redshift halos~\cite{Ralegankar:2024arh,Ralegankar:2024ekl}.

With the advent of JWST, we can now probe PMFs through the very first galaxies. JWST has revealed an overabundance of bright high-$z$ galaxies~\cite{Castellano:2025vkt,Perez-Gonzalez:2025bqr,2024ApJ...960...56H}, overmassive black holes~\cite{2023Natur.621...51D,2023ApJ...953L..29L,2023ApJ...959...39H,Bogdan:2023ilu,Maiolino:2023bpi,2024ApJ...966..176Y,2023ApJ...954L...4K}, little red dots~\cite{Yang:2023pnw,2025ApJ...986..165T,2024ApJ...964...39G,Matthee:2023utn}, and growing evidence for Pop-III stars~\cite{2023A&A...678A.173V,Visbal:2025xmd}; all connected to small-scale structure whose abundance grows with redshift. A notable consequence was the ``reionisation crisis''~\cite{Munoz:2024fas}, wherein joint analyses of the UV luminosity function and ionising photon production inferred a reionisation history in tension with both the CMB optical depth and Lyman-$\alpha$ constraints. This discrepancy has since been alleviated by relaxing the assumed local-universe relation between galaxy magnitude and UV spectral slope~\cite{Austin:2025hxd}, though leaving little room for extra ionising photon production. In this \textit{letter}, we derive constraints on PMFs by fitting the UVLF against the latest JWST data and computing the resulting reionisation history, finding $\sqrt{\left\langle B^2 \right\rangle} < 0.27\,{\rm nG}$ and $< 0.18\,{\rm nG}$ for $n_B = -2$ and $n_B = 2$ respectively at $95\%\,{\rm CL}$.

\vspace{5pt}\noindent\textbf{Matter power spectrum  -- }  We consider a power-law initial primordial magnetic field (PMF) spectrum, $P_B^{\rm lin} \propto k^{n_B}$, and focus on the cases $n_B = -2$ and $n_B = 2$ for which there are templates fitted to magnetohydrodynamic (MHD) simulations. These choices correspond to representative red and blue spectra, capturing inflationary and causal (e.g. phase transition) magnetogenesis scenarios, respectively~\cite{Durrer:2013pga}. We parametrize the amplitude of the spectrum in terms of the smoothed magnetic field at $1\,$Mpc scale, 

\be
    B_{1\,{\rm Mpc}}^{2} \equiv \int \frac{d^{3}k}{(2\pi)^{3}}\, P_B^{\rm lin}(k)\, e^{-(k/{\rm Mpc}^{-1})^2} \,.
\ee

PMFs generate post-recombination density perturbations via the Lorentz force that acts on baryons. This adds an additional small-scale contribution to the matter power spectrum beyond the standard adiabatic CDM component, modifying structure at large wavenumbers. We approximate the total matter power spectrum as 
\be \label{eq:Delta2}
    \Delta^2(k) = \Delta_{\rm CDM}^2(k) + \Delta_B^2(k) \,,
\ee
where the $\Delta_{\rm CDM}^2(k)$  is the CDM contribution (homogeneous solution) that we compute using the transfer fucntion from~\cite{Eisenstein:1997ik}. 

For the contribution sourced by PMFs, we follow Ref.~\cite{Ralegankar:2024arh} where the nonlinear regime was calibrated against MHD simulations. Assuming Gaussian, non-helical magnetic fields, the full PMF-induced component is given by  
\begin{equation} \label{eq:DeltaB}
    \Delta_B^2(k,a) =
    \begin{cases}
    \frac{\Delta_{\rm lin}^2(k,a)}{\left[1+\left( \lambda_J k \right)^{p}\right]^{(2n_B+10)/p}} , \qquad &n_B=-2,\\[8pt]
    \frac{\gamma\,(\lambda_J k)^{7}}{\left[1+(\lambda_J k)^{p}\right]^{7/p}}
    \left[\dfrac{\xi(k,a)}{\xi(k,0.01)}\right]^{2}, \,\, &n_B=2,
    \end{cases}
\end{equation}
with~\footnote{We adopt the Planck 2015 cosmological parameters, $\Omega_m = 0.308$, $\Omega_b = 4.82 \times 10^{-2}$, and $h = 0.678$.~\cite{Planck:2015fie}.}
\begin{align}
    \Delta_{\rm lin}^2(k,a) &\simeq 0.918\times 10^{-4}\,\xi^2(k,a) \\
    &\times \, \left(\frac{k}{\rm Mpc^{-1}}\right)^{2n_B+10}
    \left(\frac{B_{1\,{\rm Mpc}}}{\rm nG}\right)^4 G_{n_B} \,,
\end{align}
where $G_{n_B}$ is a numerical coefficient~\cite{Ralegankar:2024arh}. The magnetic Jeans scale is 
\begin{equation} \label{eq:kJ_def}
    \lambda_{J} = \left[ 0.1\,\kappa_J \left(\frac{B_{1\,{\rm Mpc}}}{\rm nG}\right) \right]^{\frac{2}{n_B+5}} {\rm Mpc} \,,
\end{equation}
and the simulation-calibrated parameter values are $(\kappa_J,p) = (1.17,\,1.94)$ for $n_B=-2$ and $(\kappa_J,p,\gamma) = (0.593,\,2.36,\,0.244)$ for $n_B=2$. The growth function $\xi(k,a)$ is obtained by solving the coupled perturbation equations of baryons and DM from recombination with vanishing initial conditions~\cite{Ralegankar:2024arh}. In general, the growth function deviates from the CDM one because the magnetic forces are affecting the perturbations. However, for the late-Universe regime most relevant for our study $(z<50)$, the perturbations evolve according to the CDM growth function. 

Recently, Ref.~\cite{Facchinetti:2026wed} emphasized the effect of PMF dissipation through turbulence and ambipolar diffusion on small-scale structures. We note that the simulation-calibrated power spectrum~\eqref{eq:DeltaB} includes the effect of PMF decay through turbulence, which dominates PMF energy dissipation at $z \gtrsim 100$. At later times, ambipolar diffusion becomes the primary dissipation channel. However, by this time the PMF-induced density perturbations have already been seeded and their subsequent evolution is gravity dominated, so ambipolar diffusion does not significantly affect the small-scale structures.

The above description applies when PMFs significantly affect the matter power spectrum and magnetic pressure dominates over baryonic thermal pressure. To ensure this, we require that the Jeans scale is larger than the thermal Jeans scale~\cite{Tseliakhovich:2010bj}
\begin{equation}
    \lambda_{\rm th} = \sqrt{\frac{2\Omega_m}{3\Omega_b}} \frac{c_b}{aH},
\end{equation}
where $c_b$ is the baryonic sound speed~\cite{Tseliakhovich:2010bj}. We evaluate the condition $\lambda_J \gtrsim \lambda_{\rm th}$ at $z\simeq 100$, when the thermal Jeans scale has stabilised well after recombination, which sets up a minimum $B$ field for a range of validity of our approach. For the RMS strength of the PMF, defined as
\be
    \langle B^{2}\rangle \equiv \int \frac{d^{3}k}{(2\pi)^{3}}\, P_B^{\rm lin}(k) e^{-k^2\lambda_D^2}
    %= \frac{A}{4\pi^{2}}\, \lambda_{D}^{-(n_B+3)} \Gamma\!\left(\frac{n_B+3}{2}\right)
    = B_{1\,{\rm Mpc}}^2 \left(\frac{\rm Mpc}{\lambda_D}\right)^{n_B+3} ,
\ee
with the damping scale
\be
    \lambda_D = \left[ 0.1\,\kappa(n_B) \left(\frac{B_{1\,{\rm Mpc}}}{\rm nG}\right) \right]^{\frac{2}{n_B+5}} {\rm Mpc} \,,
\ee
where the factor $\kappa_D$ is extracted from magnetohydrodynamic simulations~\cite{Ralegankar:2024arh} ($\kappa_D=0.8$ and $\kappa_D=0.9$ for $n_B = -2$ and $n_B = 2$, respectively), this corresponds to
\begin{equation} \label{eq:validity}
    \sqrt{\langle B^2 \rangle} >
    \begin{cases}
        0.09\, {\rm nG} \,, & n_B = -2 \\
        0.08\, {\rm nG} \,, & n_B = 2
    \end{cases} \,.
\end{equation}

A key connection between astrophysics and PMFs is encoded in the halo mass function $\mathrm{d}n_h/\mathrm{d}\ln M$, which we compute from the matter power spectrum~\eqref{eq:Delta2} within the excursion-set formalism formalism~\cite{Bond:1990iw} with ellipsoidal collapse~\cite{Sheth:1999su,Sheth:2001dp}. We note that PMFs also induce additional effects that impact star formation. While Lorentz forces acting on the baryon fluid enhance the baryon fraction and, consequently, star formation in primordial halos~\cite{Ralegankar:2024ekl}, PMFs also dissipate via ambipolar diffusion, which heats the gas and shifts the onset of star formation to higher halo masses and later times~\cite{Sethi:2004pe,Schleicher:2009zb}. We neglect the former effect, which makes our constraints conservative. Furthermore, while we adopt a redshift dependence of the cutoff halo mass motivated by atomic cooling, we treat its overall normalization as a free parameter and marginalize over it, thereby absorbing potential shifts in the cutoff mass induced by ambipolar diffusion heating.

\vspace{5pt}\noindent\textbf{UVLF -- } We compute the UVLF following Ref.~\cite{Ellis:2025xju}. The UVLF as a function of the UV luminosity $L$ is obtained from the halo mass function as
\be
    \Phi_{\rm UV}(L) = \int \td \tilde{L}  \, \td M \, \frac{\td P(L|\tilde{L})}{\td L} \frac{\td P(\tilde{L}|M)}{\td \tilde{L}} \frac{\td n_h}{\td M} \,.
\ee
The first distribution inside the integral captures the effects of dust attenuation and weak-lensing correction, while the second relates the luminosity of the galaxy to the mass of its host halo and encodes the SFR. We parametrize the SFR as a broken power-law with an exponential cutoff:
\be
    {\rm SFR} \propto \frac{e^{-M_t/M}}{\beta \left(M/M_c\right)^{-\alpha} + \alpha \left(M/M_c\right)^{\beta}} \,, 
\ee
and assume that the cutoff follows the redshift dependence expected for atomic cooling halos~\cite{Barkana:2000fd}, $M_t = M_{\rm cut} [(1+z)/10]^{-3/2}$. Furthermore, we accommodate enhancement due to a change in the stellar population and scatter in the SFR-luminosity relation, and allow for a high redshift reduction in the feedback effects that determine the SFR (see~\cite{Ellis:2025xju} for details).

\begin{figure}
    \centering
    \includegraphics[width=\linewidth]{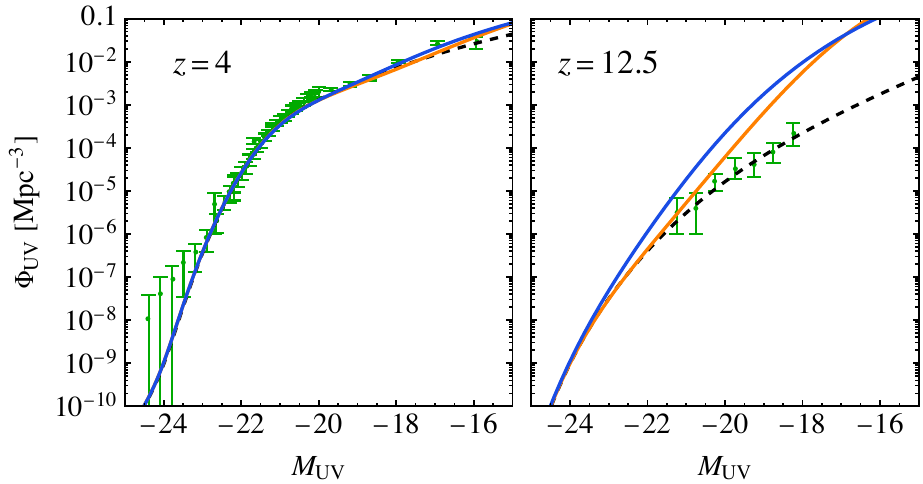}
    \vspace{-0.5 cm}
    \caption{UVLFs for CDM best fit (black dashed) and for the same SFR parameters including PMFs with spectral indices $n_B = -2$ with $B_{1\,{\rm Mpc}} = 0.5\,$nG (blue) and $n_B = 2$ with $B_{1\,{\rm Mpc}} = 0.003\,$nG (orange). The data points in the left and right panels are from HST and JWST observations, respectively.}
    \label{fig:uvlf}
\end{figure}

The effect of PMFs on the UVLF is illustrated in Fig.~\ref{fig:uvlf} as a function of the UV magnitude, defined by $\log_{10}(L/{\rm erg}) \equiv 0.4(51.63 - M_{\rm UV})$. Compared to the dashed curve, which represents the best-fit CDM case, we find that PMFs primarily enhances the high-$M_{\rm UV}$ regime, corresponding to the lowest luminosity galaxies or, equivalently, the lowest mass DM halos hosting them. This effect becomes more pronounced at higher redshifts, where JWST observations constrain the UVLF.

\begin{figure}
    \centering
    \includegraphics[width=\linewidth]{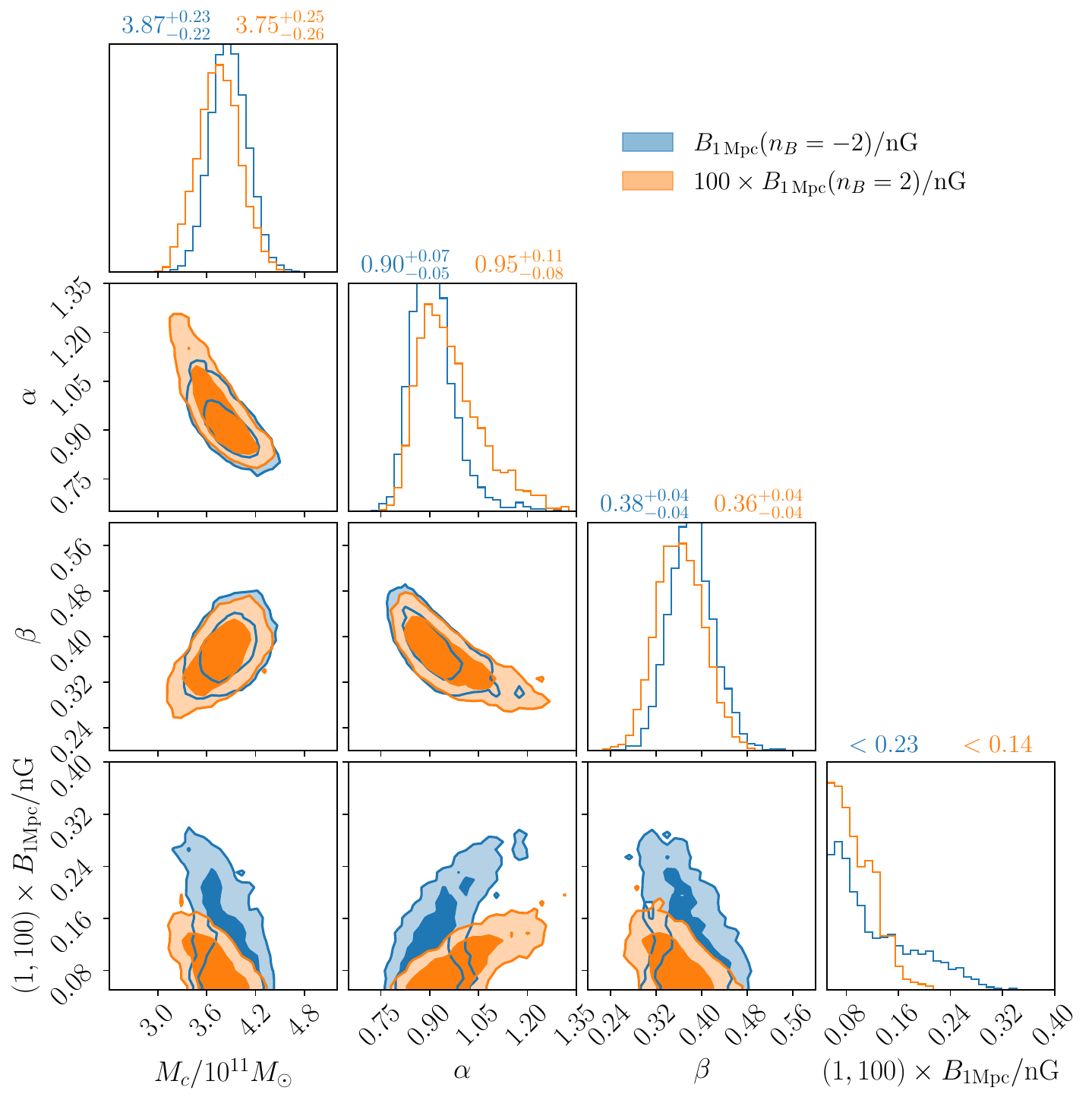}
    \caption{Posteriors from the UVLF fit. The contours show the 68\% and 95\% CL regions.}
\label{fig:corner_uvlf}
\end{figure}

We fit the SFR parametrization, together with the PMF strength $B_{1\,{\rm Mpc}}$ against data from the Hubble Space Telescope (HST)~\cite{2021AJ....162...47B,2022ApJS..259...20H} and JWST~\cite{2024MNRAS.533.3222D,Perez-Gonzalez:2025bqr,Castellano:2025vkt,Naidu:2025xfo}.\footnote{We have checked that excluding the highest redshift observations does not significantly alter the posteriors. In particular, restricting to data at $z<15$ relaxes the constraint on the PMF strength by less than 10\%.} The posteriors of the UVLF fit for the parameters that correlate the most strongly with the PMF strength are shown in Fig.~\ref{fig:corner_uvlf}. As PMFs enhance the low-luminosity part of the UVLF (see Fig.~\ref{fig:uvlf}), their effect can be compensated for by increasing $\alpha$ that controls the low mass power-law of the SFR. As seen in Fig.~\ref{fig:corner_uvlf}, achieving a good fit then requires a decrease in $\beta$ and $M_c$. However, this compensation cannot accommodate arbitrarily large deviations from the CDM limit, leading to an upper bound on the PMF strength. At 95\% credible level (CL), these bounds translate into
\be \label{eq:uvlfbound}
    \sqrt{\langle B^2\rangle} < 
    \begin{cases}
        0.87 \, {\rm nG} \,, & n_B = -2 \\
        0.85 \, {\rm nG} \,, & n_B = 2
    \end{cases} \,.
\ee
Notice that these bounds lie well within the validity regime of our computations, as given in Eq.~\eqref{eq:validity}.

\vspace{5pt}\noindent\textbf{Reionization -- }  The stellar populations that are measured through the UVLF are also the primary sources of ionizing photons responsible for reionization. The fraction of ionized hydrogen evolves as~\cite{1999ApJ...514..648M}
\be\label{eq:dotxHII}
    \dot{x}_{\rm HII}(z) = \frac{\dot{n}_{\rm ion}(z)}{n_{\rm H}(z)}-\frac{x_{\rm HII}(z)}{t_{\rm rec}(z)} \,,
\ee
where the first term describes the production of ionising photons and the second term the recapture of free electrons by the ionized hydrogen. The latter is characterised by the recombination timescale $t_{\rm rec}(z)$~\cite{2012ApJ...747..100S}. The connection of the ionised fraction evolution to the UVLF is through the comoving ionizing photon production rate density, given by 
\be \label{eq:source}
    \dot{n}_{\rm ion} = (1+z)^3 \int \td L \, \Phi_{\rm UV}(L) \dot{N}_{\rm ion}(L) f_{\rm esc}\,, 
\ee
where $\dot{N}_{\rm ion}$ is the production of photons that ionised the medium as a function of the UV luminosity and $f_{\rm esc}$ is the fraction of photons that escape the galaxy. The former is expressed as $\dot{N}_{\rm ion} = L \xi_{\rm ion}$ where $\xi_{\rm ion}$ is the ionizing efficiency that is probed directly by observing the Balmer lines of galaxies.  Inspired by the results of~\cite{2025A&A...698A.302L,Papovich:2025swd}, we consider a fit of the form
\be
    \log_{10}\frac{\xi_{\rm ion}}{\rm Hz\,erg^{-1}} = a M_{\rm UV} + b + d z \,,
\ee
with $a=0.15\pm0.008$, $b=28.17 \pm 0.17$ and $d=0.05\pm0.02$ and cap the values of $\xi_{\rm ion}$ at $M_{\rm UV}>-16$ and $z<9$ to avoid extrapolation. In contrast, observational estimates of $f_{\rm esc}$ show significant discrepancies~\cite{Papovich:2025swd,2025arXiv250701096G}. For simplicity, we assume that the escape fraction $f_{\rm esc}$ is independent of $L$ and $z$, and consider a prior range of $0.05 < f_{\rm esc} < 0.5$. We set the initial condition $x_{\rm HII} = 0$ at redshit $z_*$, and consider $z_{*}\in[25\,,50]$. 

\begin{figure}
    \centering
    \includegraphics[width=0.9\linewidth]{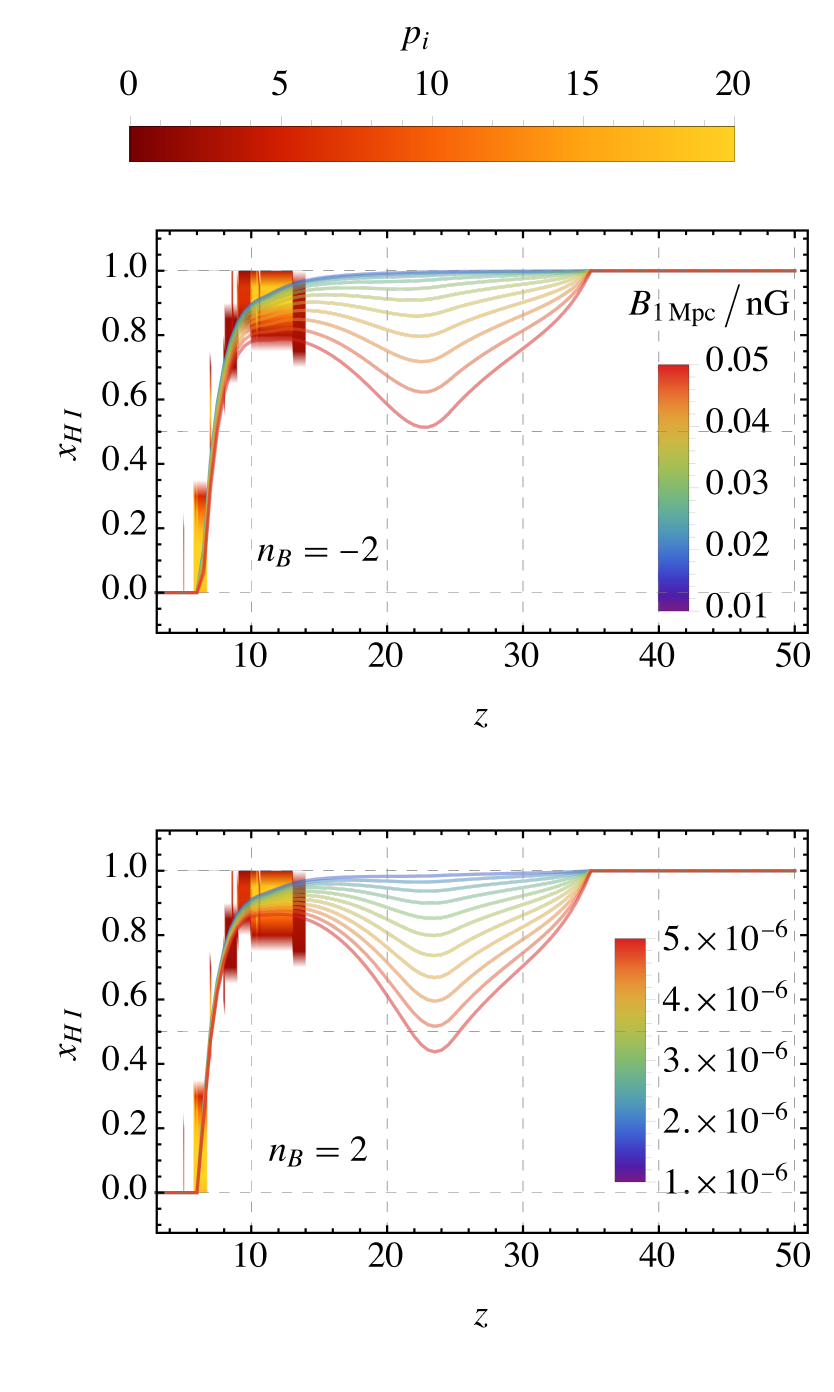}
    \vspace{-0.5 cm}
    \caption{Reionisation histories varying the PMF strength while keeping the rest of the parameters fixed to the best fit values, the data from reionisation is taken from~\cite{2025ApJS..278...33K,2025arXiv250404683U,2024ApJ...967...28N,Greig:2024atz,2023ApJ...949L..40B,Mason:2017eqr,2020MNRAS.495.3602W}.}
    \label{fig:reionisation_histories}
\end{figure}

The reionisation history is sensitive to weak PMFs that do not significantly modify the UVLF. We therefore fix the SFR parameters to their best-fit CDM values. However, for the cutoff mass $M_{\rm cut}$, the UVLF observations provide only an upper bound 
\be
    M_{\rm cut} < 3.5\times 10^8\,M_\odot
\ee
at 95\% CL, while the reionization history is quite sensitive to it. Therefore, in our reionization fit, we vary $M_{\rm cut}$ using the UVLF bound as a prior. 

The effect of PMFs on the reionisation history is illustrated in Fig.~\ref{fig:reionisation_histories}, which shows the fraction of neutral hydrogen $x_{\rm HI} = 1-x_{\rm HII}$ as a function of redshift. For $z<12$, the curves showing different PMF strengths match the Lyman-$\alpha$ observations~\cite{2025ApJS..278...33K,2025arXiv250404683U,2024ApJ...967...28N,Greig:2024atz,2023ApJ...949L..40B,Mason:2017eqr,2020MNRAS.495.3602W}, while the impact of the PMF is primarily reflected in the behaviour at higher redshifts. As we move to higher values of the magnetic field, the enhanced small-scale power leads to increased production of ionising photons, giving rise to the characteristic double reionisation history, with a dip around $z\approx 24$. This could be constrained by the Zeldovich effect~\cite{2021ApJ...908..199R}, but we instead constrain it using optical depth measurements. 

Earlier reionisation leads to a higher column density of free electrons available to scatter photons. This, in turn, results in a higher optical depth, computed by integrating along the line of sight:
\be
    \tau(z) = \int_0^{t(z)}\td t\, \sigma_{\rm T} n_e(t) \,,
\ee
where $n_e = f_e x_{\rm HII} n_{\rm H}$ is the average proper electron number density and $\sigma_{\rm T}$ is the Thompson scattering cross section. The fraction of free electrons per ionized hydrogen atom, $f_e$, is given in~\cite{2013ApJ...768...71R}. There are strict constraints that come mainly from the CMB polarisation measurements. Since this is the main quantity that constrains this double reionisation, we consider two different measurements: The first one is the most stringent Planck posterior $\tau = 0.054\pm0.0073$~\cite{Planck:2015fie}, while the second comes from the independent analysis of using only high-$l$ polarisation and prefers a slightly higher value $\tau = 0.076\pm0.015$~\cite{Giare:2023ejv} We use these as priors in our analysis.

\begin{figure}
    \centering
    \includegraphics[width=0.8\linewidth]{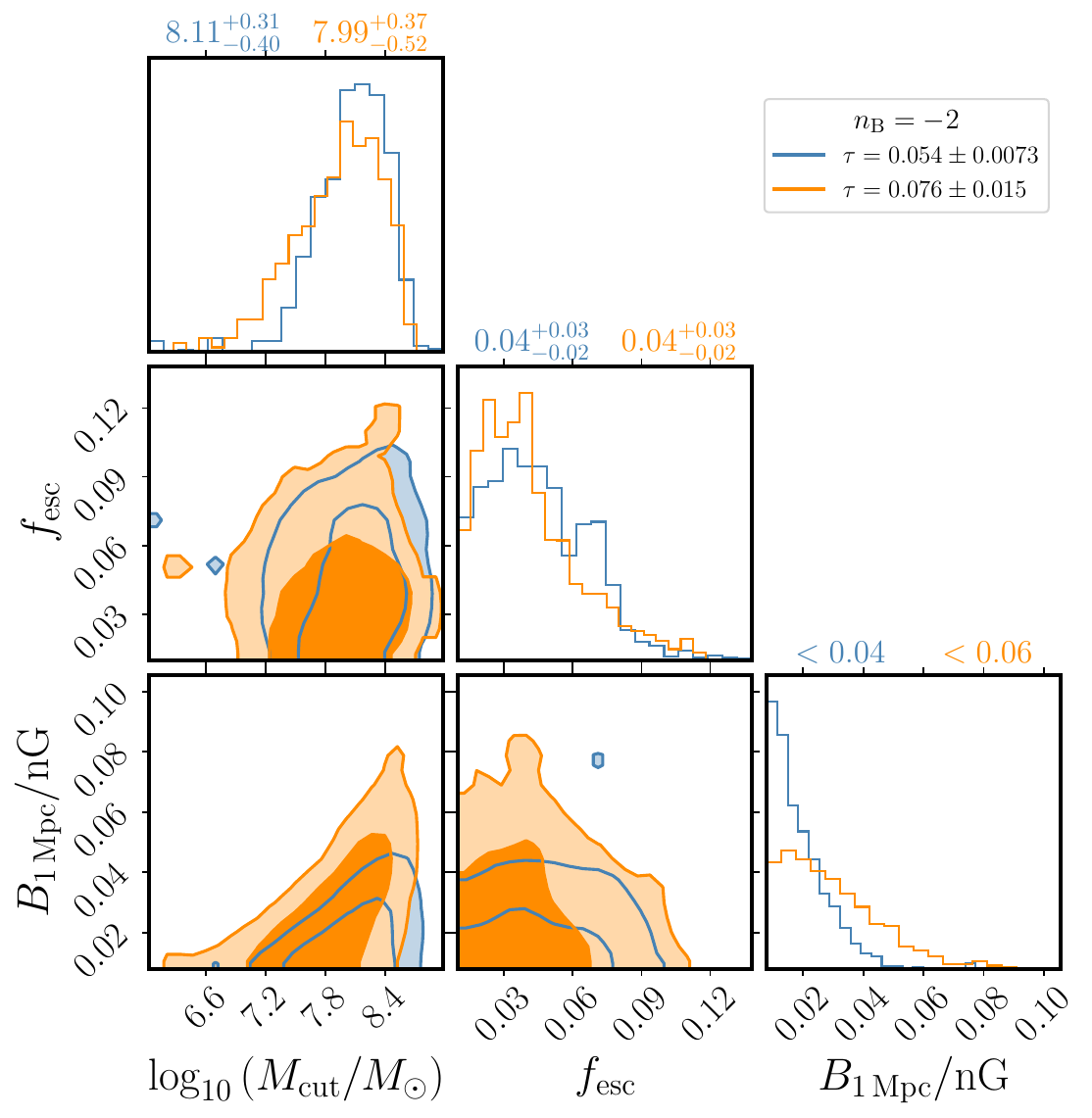}
    \caption{Posteriors the reionization fit for $n_{\rm B} = -2$. The contours show the 68\% and 95\% CL regions. The two sets of posteriors correspond to different priors for the optical depth, using Planck~\cite{Planck:2015fie} and only high-$l$ multiples~\cite{Giare:2023ejv} that prefers a higher $\tau$.}
    \label{fig:corner_inflation}
\end{figure}

We fit our reionisation model, characterised by the parameters $\vect{\theta} = (a,b,d,f_{\rm esc},M_{\rm cut},z_{*},B_{\rm 1\, Mpc})$ to the Lyman-$\alpha$ data~\cite{2025ApJS..278...33K,2025arXiv250404683U,2024ApJ...967...28N,Greig:2024atz,2023ApJ...949L..40B,Mason:2017eqr,2020MNRAS.495.3602W}. The likelihood is
\be
    \mathcal{L}(\mathcal{\vect{\theta}}) \propto P(\vect{\theta}) 
    \prod_i p_i\!\left(\langle x_{\rm HI}(\vect{\theta},z)\rangle_i \right) \,,
\ee
where $i$ labels the measurements, $p_i(x_{\rm HI})$ denotes the posterior of the ionized fraction, and the average is computed over the redshift range of the posterior.

\begin{figure}
    \centering
    \includegraphics[width=0.8\linewidth]{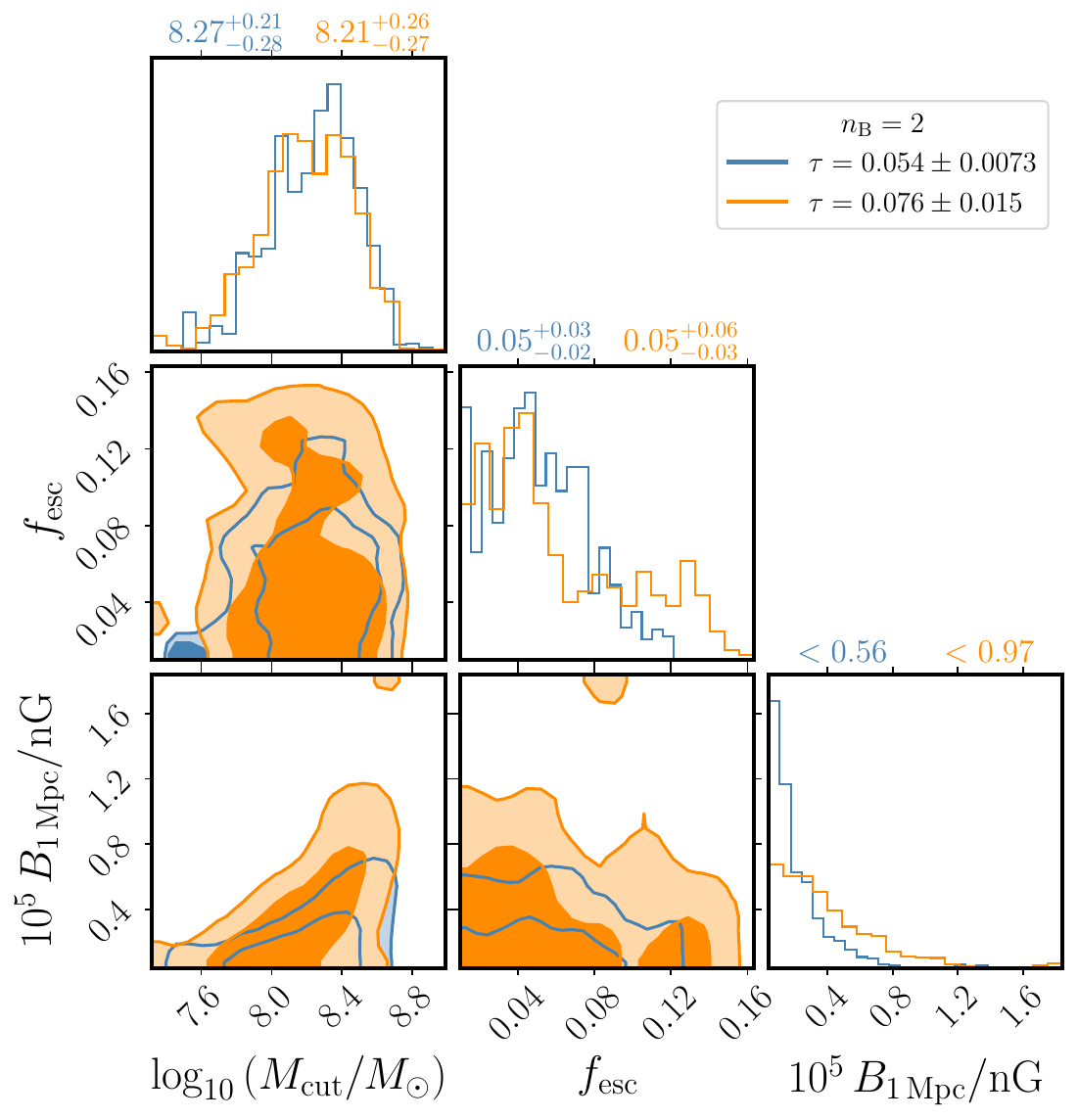}
    \caption{Same as Fig.\ref{fig:corner_inflation} for $n_{\rm B}=2$.}
    \label{fig:corner_PT}
\end{figure}

The posteriors of the reionisation history fit for $f_{\rm esc}$, $B_{\rm 1{\rm Mpc}}$, and $M_{\rm cut}$ are shown in Fig.~\ref{fig:corner_inflation} for $n_{\rm B}=-2$ and in Fig.~\ref{fig:corner_PT} for $n_{\rm B}=2$. The constraints on $(a,b,d)$ are largely prior-driven, and $z_*$ remains essentially unconstrained. In both cases, we see that the minimum halo mass for star formation, $M_{\rm cut}$ is correlated with the PMF strength: larger minimum halo masses yield fewer ionising photons, thereby permitting higher primordial magnetic field strengths. Moreover, the fraction of photons that escape galaxies and ionise the intergalactic medium, $f_{\rm esc}$, is bounded from above and correlates with the PMF strength, stronger PMFs are only compatible with smaller escape fractions. As expected, constraints on the PMFs obtained using Planck priors on the optical depth are more stringent than those derived from high-$\ell$ polarisation modes alone, though the difference is modest. The bounds obtained using the Planck priors on $\tau$ are 
\be \label{eq:reionisationbound}
    \sqrt{\langle B^2\rangle} < 
    \begin{cases}
        0.27 \, {\rm nG} \,, & n_B = -2 \\
        0.18 \, {\rm nG} \,, & n_B = 2
    \end{cases} \,.
\ee
These constraints are significantly stronger than those obtained from the UVLF observations~\eqref{eq:uvlfbound}, while still remaining within the validity range~\eqref{eq:validity} of our computations.

\vspace{5pt}\noindent\textbf{Conclusions  -- } In this letter, we have derived new constraints on primordial magnetic fields (PMFs) using two complementary high-redshift observables: the UV luminosity function (UVLF) and the reionisation history. By combining the latest JWST data with a simulation-calibrated treatment of the PMF-induced matter power spectrum, we have demonstrated that early galaxy observations provide a powerful probe of PMFs.

From the UVLF analysis, we find that PMFs are constrained at 95\% CL to $\sqrt{\left\langle B^2 \right\rangle} < 0.87\,{\rm nG}$ and $< 0.85\,{\rm nG}$ for $n_B = -2$ and $n_B = 2$, respectively. The key mechanism driving these bounds is the PMF-induced enhancement of low-luminosity galaxy counts, which cannot be fully compensated by adjustments to the star formation rate parameters. Crucially, while PMFs enhance the abundance of low-mass halos, this excess is not required by the data and, as shown in~\cite{Ellis:2025xju}, the JWST observations of the UVLF are well described within the standard CDM framework once the SFR model accommodates for enhancement due to a change in the stellar population.

The reionisation history yields considerably stronger constraints: $\sqrt{\left\langle B^2 \right\rangle} < 0.27\,{\rm nG}$ and $< 0.18\,{\rm nG}$ for $n_B = -2$ and $n_B = 2$, respectively, using Planck optical depth priors. This improvement arises because strong PMFs induce a characteristic double reionisation signature at high redshift that is incompatible with CMB constraints on optical depth $\tau$. Such signature is not indicated by the data, but represents an unavoidable consequence of enhanced small-scale structure that the observations rule out.

Lyman-$\alpha$ forest observations provide a complementary probe of PMF-induced small-scale matter fluctuations. The constraint $\sqrt{\langle B^2 \rangle} < 0.3\,$nG obtained in the nearly scale-invariant limit in~\cite{Pavicevic:2025gqi} is comparable to our constraints from reionization. Our constraints are also comparable to the CMB constraints on PMFs arising from their impact on the post-recombination ionisation history~\cite{Paoletti:2022gsn}, that give $\sqrt{\langle B^2\rangle}\lesssim 0.79\,$nG for $n_B=-2$ and $\lesssim 0.06\,$nG for $n_B=2$, and lie well above the lower limit set from blazar observations from H.E.S.S./Fermi-LAT~\cite{HESS:2023zwb}, $\langle B^2 \rangle \gtrsim 10^{-6} \,$nG for coherence lengths $ \gtrsim 0.1\,$Mpc. Future blazar observations with CTA are expected to extend this reach up to $B \simeq 10^{-3}\,$nG~\cite{Korochkin:2020pvg}, although still remaining well below our upper limits.

Our results show that the high-redshift universe provides a powerful laboratory for constraining the magnetised early cosmos.

\vspace{5pt}\noindent\emph{Acknowledgments --}  The authors thank Pranjal Ralegankar for useful discussions. The work of MOOR and MF was supported by the United Kingdom STFC Grant ST/T00679X/1. The work of JU and VV was supported by the Estonian Research Council grants TARISTU24-TK3, TARISTU24-TK10, and the Centre of Excellence programme TK202 of the Estonian Ministry of Education and Research. The work of VV was partially supported by the European Union's Horizon Europe research and innovation program under the Marie Sk\l{}odowska-Curie grant agreement No. 101065736. MOOR also wants to acknowledge support of the grant PID2024-161668NB-I00.

\bibliographystyle{JHEP.bst}
\bibliography{references.bib}

@article{Durrer:2013pga,
    author = "Durrer, Ruth and Neronov, Andrii",
    title = "{Cosmological Magnetic Fields: Their Generation, Evolution and Observation}",
    eprint = "1303.7121",
    archivePrefix = "arXiv",
    primaryClass = "astro-ph.CO",
    doi = "10.1007/s00159-013-0062-7",
    journal = "Astron. Astrophys. Rev.",
    volume = "21",
    pages = "62",
    year = "2013"
}

@article{Facchinetti:2026wed,
    author = "Facchinetti, Ga{\'e}tan and Korochkin, Alexander and Lopez-Honorez, Laura and Schwagereit, Justus",
    title = "{Reionization, UV Luminosity and 21$\,$cm Sensitivity to Primordial Magnetic Fields: Impact of Energy Losses}",
    eprint = "2604.22703",
    archivePrefix = "arXiv",
    primaryClass = "astro-ph.CO",
    month = "4",
    year = "2026"
}

@article{Pavicevic:2025gqi,
    author = "Pavi{\v{c}}evi{\'c}, Mak and Ir{\v{s}}i{\v{c}}, Vid and Viel, Matteo and Bolton, James S. and Haehnelt, Martin G. and Martin-Alvarez, Sergio and Puchwein, Ewald and Ralegankar, Pranjal",
    title = "{Constraints on Primordial Magnetic Fields from the Lyman-{\ensuremath{\alpha}} Forest}",
    eprint = "2501.06299",
    archivePrefix = "arXiv",
    primaryClass = "astro-ph.CO",
    doi = "10.1103/77rd-vkpz",
    journal = "Phys. Rev. Lett.",
    volume = "135",
    number = "7",
    pages = "071001",
    year = "2025"
}

@article{Schleicher:2009zb,
    author = "Schleicher, Dominik R. G. and Galli, Daniele and Glover, Simon C. O. and Banerjee, Robi and Palla, Francesco and Schneider, Raffaella and Klessen, Ralf S.",
    title = "{The influence of magnetic fields on the thermodynamics of primordial star formation}",
    eprint = "0904.3970",
    archivePrefix = "arXiv",
    primaryClass = "astro-ph.CO",
    doi = "10.1088/0004-637X/703/1/1096",
    journal = "Astrophys. J.",
    volume = "703",
    pages = "1096--1106",
    year = "2009"
}

@article{Sethi:2004pe,
    author = "Sethi, Shiv K. and Subramanian, Kandaswamy",
    title = "{Primordial magnetic fields in the post-recombination era and early reionization}",
    eprint = "astro-ph/0405413",
    archivePrefix = "arXiv",
    doi = "10.1111/j.1365-2966.2004.08520.x",
    journal = "Mon. Not. Roy. Astron. Soc.",
    volume = "356",
    pages = "778--788",
    year = "2005"
}

@article{Barkana:2000fd,
    author = "Barkana, Rennan and Loeb, Abraham",
    title = "{In the beginning: The First sources of light and the reionization of the Universe}",
    eprint = "astro-ph/0010468",
    archivePrefix = "arXiv",
    doi = "10.1016/S0370-1573(01)00019-9",
    journal = "Phys. Rept.",
    volume = "349",
    pages = "125--238",
    year = "2001"
}

@article{Planck:2015fie,
    author = "Ade, P. A. R. and others",
    collaboration = "Planck",
    title = "{Planck 2015 results. XIII. Cosmological parameters}",
    eprint = "1502.01589",
    archivePrefix = "arXiv",
    primaryClass = "astro-ph.CO",
    doi = "10.1051/0004-6361/201525830",
    journal = "Astron. Astrophys.",
    volume = "594",
    pages = "A13",
    year = "2016"
}

@ARTICLE{2021AJ....162...47B,
       author = {{Bouwens}, R.~J. and {Oesch}, P.~A. and {Stefanon}, M. and {Illingworth}, G. and {Labb{\'e}}, I. and {Reddy}, N. and {Atek}, H. and {Montes}, M. and {Naidu}, R. and {Nanayakkara}, T. and {Nelson}, E. and {Wilkins}, S.},
        title = "{New Determinations of the UV Luminosity Functions from z   9 to 2 Show a Remarkable Consistency with Halo Growth and a Constant Star Formation Efficiency}",
      journal = {\aj},
         year = 2021,
        month = aug,
       volume = {162},
       number = {2},
          eid = {47},
        pages = {47},
          doi = {10.3847/1538-3881/abf83e},
archivePrefix = {arXiv},
       eprint = {2102.07775},
 primaryClass = {astro-ph.GA},
       adsurl = {https://ui.adsabs.harvard.edu/abs/2021AJ....162...47B}
}

@ARTICLE{2022ApJS..259...20H,
       author = {{Harikane}, Yuichi and {Ono}, Yoshiaki and {Ouchi}, Masami and {Liu}, Chengze and {Sawicki}, Marcin and {Shibuya}, Takatoshi and {Behroozi}, Peter S. and {He}, Wanqiu and {Shimasaku}, Kazuhiro and {Arnouts}, Stephane and {Coupon}, Jean and {Fujimoto}, Seiji and {Gwyn}, Stephen and {Huang}, Jiasheng and {Inoue}, Akio K. and {Kashikawa}, Nobunari and {Komiyama}, Yutaka and {Matsuoka}, Yoshiki and {Willott}, Chris J.},
        title = "{GOLDRUSH. IV. Luminosity Functions and Clustering Revealed with  4,000,000 Galaxies at z   2-7: Galaxy-AGN Transition, Star Formation Efficiency, and Implication for Evolution at z > 10}",
      journal = {\apjs},
         year = 2022,
        month = mar,
       volume = {259},
       number = {1},
          eid = {20},
        pages = {20},
          doi = {10.3847/1538-4365/ac3dfc},
archivePrefix = {arXiv},
       eprint = {2108.01090},
 primaryClass = {astro-ph.GA},
       adsurl = {https://ui.adsabs.harvard.edu/abs/2022ApJS..259...20H}
}

@ARTICLE{2024MNRAS.533.3222D,
       author = {{Donnan}, C.~T. and {McLure}, R.~J. and {Dunlop}, J.~S. and {McLeod}, D.~J. and {Magee}, D. and {Arellano-C{\'o}rdova}, K.~Z. and {Barrufet}, L. and {Begley}, R. and {Bowler}, R.~A.~A. and {Carnall}, A.~C. and {Cullen}, F. and {Ellis}, R.~S. and {Fontana}, A. and {Illingworth}, G.~D. and {Grogin}, N.~A. and {Hamadouche}, M.~L. and {Koekemoer}, A.~M. and {Liu}, F.-Y. and {Mason}, C. and {Santini}, P. and {Stanton}, T.~M.},
        title = "{JWST PRIMER: a new multifield determination of the evolving galaxy UV luminosity function at redshifts $z \simeq 9 - 15$}",
      journal = {\mnras},
         year = 2024,
        month = sep,
       volume = {533},
       number = {3},
        pages = {3222-3237},
          doi = {10.1093/mnras/stae2037},
archivePrefix = {arXiv},
       eprint = {2403.03171},
 primaryClass = {astro-ph.GA},
       adsurl = {https://ui.adsabs.harvard.edu/abs/2024MNRAS.533.3222D}
}

@article{Perez-Gonzalez:2025bqr,
    author = "P{\'e}rez-Gonz{\'a}lez, Pablo G. and others",
    title = "{{The Rise of the Galactic Empire: Ultraviolet Luminosity Functions at $z \sim 17$ and $z \sim 25$ Estimated with the MIDIS+NGDEEP Ultra-deep JWST/NIRCam Data Set}}",
    eprint = "2503.15594",
    archivePrefix = "arXiv",
    primaryClass = "astro-ph.GA",
    doi = "10.3847/1538-4357/adf8c9",
    journal = "Astrophys. J.",
    volume = "991",
    number = "2",
    pages = "179",
    year = "2025"
}

@article{Castellano:2025vkt,
    author = "Castellano, M. and others",
    title = "{{Pushing JWST to the extremes: Search and scrutiny of bright galaxy candidates at $z \simeq 15 - 30$}}",
    eprint = "2504.05893",
    archivePrefix = "arXiv",
    primaryClass = "astro-ph.GA",
    doi = "10.1051/0004-6361/202555082",
    journal = "Astron. Astrophys.",
    volume = "704",
    pages = "A158",
    year = "2025"
}

@article{Naidu:2025xfo,
    author = "Naidu, Rohan P. and others",
    title = "{A Cosmic Miracle: A Remarkably Luminous Galaxy at $z_{\rm{spec}}=14.44$ Confirmed with JWST}",
    eprint = "2505.11263",
    archivePrefix = "arXiv",
    primaryClass = "astro-ph.GA",
    month = "5",
    year = "2025"
}

@article{Eisenstein:1997ik,
    author = "Eisenstein, Daniel J. and Hu, Wayne",
    title = "{Baryonic features in the matter transfer function}",
    eprint = "astro-ph/9709112",
    archivePrefix = "arXiv",
    reportNumber = "IASSNS-AST-97-51",
    doi = "10.1086/305424",
    journal = "Astrophys. J.",
    volume = "496",
    pages = "605",
    year = "1998"
}

@article{Bond:1990iw,
    author = "Bond, J. R. and Cole, S. and Efstathiou, G. and Kaiser, Nick",
    title = "{Excursion set mass functions for hierarchical Gaussian fluctuations}",
    reportNumber = "CFPA-TH-90-015",
    doi = "10.1086/170520",
    journal = "Astrophys. J.",
    volume = "379",
    pages = "440",
    year = "1991"
}

@article{Sheth:1999su,
    author = "Sheth, Ravi K. and Mo, H. J. and Tormen, Giuseppe",
    title = "{Ellipsoidal collapse and an improved model for the number and spatial distribution of dark matter haloes}",
    eprint = "astro-ph/9907024",
    archivePrefix = "arXiv",
    doi = "10.1046/j.1365-8711.2001.04006.x",
    journal = "Mon. Not. Roy. Astron. Soc.",
    volume = "323",
    pages = "1",
    year = "2001"
}

@article{Sheth:2001dp,
    author = "Sheth, Ravi K. and Tormen, Giuseppe",
    title = "{An Excursion Set Model of Hierarchical Clustering : Ellipsoidal Collapse and the Moving Barrier}",
    eprint = "astro-ph/0105113",
    archivePrefix = "arXiv",
    reportNumber = "FERMILAB-PUB-01-061-A",
    doi = "10.1046/j.1365-8711.2002.04950.x",
    journal = "Mon. Not. Roy. Astron. Soc.",
    volume = "329",
    pages = "61",
    year = "2002"
}

@inbook{Beck:2013bxa,
    author = "Beck, Rainer and Wielebinski, Richard",
    title = "{Magnetic Fields in the Milky Way and in Galaxies}",
    eprint = "1302.5663",
    archivePrefix = "arXiv",
    primaryClass = "astro-ph.GA",
    doi = "10.1007/978-94-007-5612-0_13",
    month = "2",
    year = "2013"
}

@article{Reiners:2012bb,
    author = "Reiners, Ansgar",
    title = "{Observations of Cool-Star Magnetic Fields}",
    eprint = "1203.0241",
    archivePrefix = "arXiv",
    primaryClass = "astro-ph.SR",
    doi = "10.12942/lrsp-2012-1",
    journal = "Living Rev. Sol. Phys.",
    volume = "9",
    pages = "1",
    year = "2012"
}

@article{Beck:2008ty,
    author = "Beck, Rainer",
    editor = "Aharonian, Felix A. and Hofmann, Werner and Rieger, Frank",
    title = "{Galactic and Extragalactic Magnetic Fields}",
    eprint = "0810.2923",
    archivePrefix = "arXiv",
    primaryClass = "astro-ph",
    doi = "10.1063/1.3076806",
    journal = "AIP Conf. Proc.",
    volume = "1085",
    number = "1",
    pages = "83--96",
    year = "2009"
}

@ARTICLE{2010SSRv..152..651S,
       author = {{Stevenson}, David J.},
        title = "{Planetary Magnetic Fields: Achievements and Prospects}",
      journal = {Space Sci. Rev.},
         year = "2010",
        month = "May",
       volume = {152},
       number = {1-4},
        pages = {651-664},
          doi = {10.1007/s11214-009-9572-z},
       adsurl = {https://ui.adsabs.harvard.edu/abs/2010SSRv..152..651S}
}

@article{Govoni_2019,
   title={A radio ridge connecting two galaxy clusters in a filament of the cosmic web},
   volume={364},
   ISSN={1095-9203},
   url={http://dx.doi.org/10.1126/science.aat7500},
   DOI={10.1126/science.aat7500},
   number={6444},
   journal={Science},
   publisher={American Association for the Advancement of Science (AAAS)},
   author={Govoni, F. and Orrù, E. and Bonafede, A. and Iacobelli, M. and Paladino, R. and Vazza, F. and Murgia, M. and Vacca, V. and Giovannini, G. and Feretti, L. and Loi, F. and Bernardi, G. and Ferrari, C. and Pizzo, R. F. and Gheller, C. and Manti, S. and Brüggen, M. and Brunetti, G. and Cassano, R. and de Gasperin, F. and Enßlin, T. A. and Hoeft, M. and Horellou, C. and Junklewitz, H. and Röttgering, H. J. A. and Scaife, A. M. M. and Shimwell, T. W. and van Weeren, R. J. and Wise, M.},
   year={2019},
   month=jun, pages={981–984} }

@article{Vogt:2005xf,
    author = "Vogt, Corina and Ensslin, Torsten A.",
    title = "{A Bayesian view on Faraday rotation maps - Seeing the magnetic power spectra in galaxy clusters}",
    eprint = "astro-ph/0501211",
    archivePrefix = "arXiv",
    doi = "10.1051/0004-6361:20041839",
    journal = "Astron. Astrophys.",
    volume = "434",
    pages = "67",
    year = "2005"
}

@article{Osinga:2022tos,
    author = {Osinga, E. and van Weeren, R. J. and Andrade-Santos, F. and Rudnick, L. and Bonafede, A. and Clarke, T. and Duncan, K. and Giacintucci, S. and Mroczkowski, Tony and R{\"o}ttgering, H. J. A.},
    title = "{The detection of cluster magnetic fields via radio source depolarisation}",
    eprint = "2207.09717",
    archivePrefix = "arXiv",
    primaryClass = "astro-ph.CO",
    doi = "10.1051/0004-6361/202243526",
    journal = "Astron. Astrophys.",
    volume = "665",
    pages = "A71",
    year = "2022"
}

@article{McBride:2012mm,
    author = "McBride, James and Heiles, Carl",
    title = "{An Arecibo Survey for Zeeman Splitting in OH Megamaser Galaxies}",
    eprint = "1211.2023",
    archivePrefix = "arXiv",
    primaryClass = "astro-ph.GA",
    doi = "10.1088/0004-637X/763/1/8",
    journal = "Astrophys. J.",
    volume = "763",
    pages = "8",
    year = "2013"
}

@article{Robishaw:2008ti,
    author = "Robishaw, Timothy and Quataert, Eliot and Heiles, Carl",
    title = "{Extragalactic Zeeman Detections in OH Megamasers}",
    eprint = "0803.1832",
    archivePrefix = "arXiv",
    primaryClass = "astro-ph",
    doi = "10.1086/588031",
    journal = "Astrophys. J.",
    volume = "680",
    pages = "981",
    year = "2008"
}

@article{Geach:2023enq,
    author = "Geach, J. E. and Lopez-Rodriguez, E. and Doherty, M. J. and Chen, Jianhang and Ivison, R. J. and Bendo, G. J. and Dye, S. and Coppin, K. E. K.",
    title = "{Polarized thermal emission from dust in a galaxy at redshift 2.6}",
    eprint = "2309.02034",
    archivePrefix = "arXiv",
    primaryClass = "astro-ph.GA",
    doi = "10.1038/s41586-023-06346-4",
    journal = "Nature",
    volume = "621",
    number = "7979",
    pages = "483--486",
    year = "2023"
}

@article{Bernet:2008qp,
    author = "Bernet, Martin L. and Miniati, Francesco and Lilly, Simon J. and Kronberg, Philipp P. and Dessauges-Zavadsky, Miroslava",
    title = "{Strong magnetic fields in normal galaxies at high redshifts}",
    eprint = "0807.3347",
    archivePrefix = "arXiv",
    primaryClass = "astro-ph",
    doi = "10.1038/nature07105",
    journal = "Nature",
    volume = "454",
    pages = "302--304",
    year = "2008"
}

@article{Martin-Alvarez:2021jsh,
    author = "Martin-Alvarez, Sergio and Devriendt, Julien and Slyz, Adrianne and Sijacki, Debora and Richardson, Mark L. A. and Katz, Harley",
    title = "{Towards convergence of turbulent dynamo amplification in cosmological simulations of galaxies}",
    eprint = "2111.06901",
    archivePrefix = "arXiv",
    primaryClass = "astro-ph.GA",
    doi = "10.1093/mnras/stac1099",
    journal = "Mon. Not. Roy. Astron. Soc.",
    volume = "513",
    number = "3",
    pages = "3326--3344",
    year = "2022"
}

@article{Schober:2013aoa,
    author = "Schober, Jennifer and Schleicher, Dominik R. G. and Klessen, Ralf S.",
    title = "{Magnetic Field Amplification in Young Galaxies}",
    eprint = "1310.0853",
    archivePrefix = "arXiv",
    primaryClass = "astro-ph.GA",
    doi = "10.1051/0004-6361/201322185",
    journal = "Astron. Astrophys.",
    volume = "560",
    pages = "A87",
    year = "2013"
}

@article{Hanayama:2005hd,
    author = "Hanayama, Hidekazu and Takahashi, Keitaro and Kotake, Kei and Oguri, Masamune and Ichiki, Kiyotomo and Ohno, Hiroshi",
    title = "{Biermann mechanism in primordial supernova remnant and seed magnetic fields}",
    eprint = "astro-ph/0501538",
    archivePrefix = "arXiv",
    doi = "10.1086/491575",
    journal = "Astrophys. J.",
    volume = "633",
    pages = "941",
    year = "2005"
}

@article{Sur:2012zj,
    author = "Sur, Sharanya and Federrath, Christoph and Schleicher, Dominik R. G. and Banerjee, Robi and Klessen, Ralf S.",
    title = "{Magnetic field amplification during gravitational collapse - Influence of initial conditions on dynamo evolution and saturation}",
    eprint = "1202.3206",
    archivePrefix = "arXiv",
    primaryClass = "astro-ph.SR",
    doi = "10.1111/j.1365-2966.2012.21100.x",
    journal = "Mon. Not. Roy. Astron. Soc.",
    volume = "423",
    pages = "3148",
    year = "2012"
}

@ARTICLE{1950ZNatA...5...65B,
       author = {{Biermann}, L.},
        title = "{{\"U}ber den Ursprung der Magnetfelder auf Sternen und im interstellaren Raum (miteinem Anhang von A. Schl{\"u}ter)}",
      journal = {Zeitschrift Naturforschung Teil A},
         year = 1950,
        month = jan,
       volume = {5},
        pages = {65},
       adsurl = {https://ui.adsabs.harvard.edu/abs/1950ZNatA...5...65B}
}

@article{Donnert:2008sn,
    author = "Donnert, J. and Dolag, K. and Lesch, H. and Muller, E.",
    title = "{Cluster Magnetic Fields from Galactic Outflows}",
    eprint = "0808.0919",
    archivePrefix = "arXiv",
    primaryClass = "astro-ph",
    doi = "10.1111/j.1365-2966.2008.14132.x",
    journal = "Mon. Not. Roy. Astron. Soc.",
    volume = "392",
    pages = "1008--1021",
    year = "2009"
}

@article{Vazza:2014jga,
    author = {Vazza, F. and Br{\"u}ggen, M. and Gheller, C. and Wang, P.},
    title = "{On the amplification of magnetic fields in cosmic filaments and galaxy clusters}",
    eprint = "1409.2640",
    archivePrefix = "arXiv",
    primaryClass = "astro-ph.CO",
    doi = "10.1093/mnras/stu1896",
    journal = "Mon. Not. Roy. Astron. Soc.",
    volume = "445",
    number = "4",
    pages = "3706--3722",
    year = "2014"
}

@article{Neronov:2010gir,
    author = "Neronov, A. and Vovk, I.",
    title = "{Evidence for strong extragalactic magnetic fields from Fermi observations of TeV blazars}",
    eprint = "1006.3504",
    archivePrefix = "arXiv",
    primaryClass = "astro-ph.HE",
    doi = "10.1126/science.1184192",
    journal = "Science",
    volume = "328",
    pages = "73--75",
    year = "2010"
}

@article{Broderick:2011av,
    author = "Broderick, Avery E. and Chang, Philip and Pfrommer, Christoph",
    title = "{The Cosmological Impact of Luminous TeV Blazars I: Implications of Plasma Instabilities for the Intergalactic Magnetic Field and Extragalactic Gamma-Ray Background}",
    eprint = "1106.5494",
    archivePrefix = "arXiv",
    primaryClass = "astro-ph.CO",
    doi = "10.1088/0004-637X/752/1/22",
    journal = "Astrophys. J.",
    volume = "752",
    pages = "22",
    year = "2012"
}

@ARTICLE{1978ApJ...224..337W,
       author = {{Wasserman}, I.},
        title = "{On the origins of galaxies, galactic angular momenta, and galactic magnetic fields.}",
      journal = {\apj},
         year = 1978,
        month = sep,
       volume = {224},
        pages = {337-343},
          doi = {10.1086/156381},
       adsurl = {https://ui.adsabs.harvard.edu/abs/1978ApJ...224..337W}
}

@ARTICLE{2003JApA...24...51G,
       author = {{Gopal}, Rajesh and {Sethi}, Shiv K.},
        title = "{Large Scale Magnetic Fields: Density Power Spectrum in Redshift Space}",
      journal = {Journal of Astrophysics and Astronomy},
         year = 2003,
        month = sep,
       volume = {24},
       number = {3-4},
        pages = {51-67},
          doi = {10.1007/BF02702312},
       adsurl = {https://ui.adsabs.harvard.edu/abs/2003JApA...24...51G}
}

@article{Kim:1994zh,
    author = "Kim, Eun-jin and Olinto, Angela and Rosner, Robert",
    title = "{Generation of density perturbations by primordial magnetic fields}",
    eprint = "astro-ph/9412070",
    archivePrefix = "arXiv",
    reportNumber = "FERMILAB-PUB-94-428-A",
    doi = "10.1086/177667",
    journal = "Astrophys. J.",
    volume = "468",
    pages = "28",
    year = "1996"
}

@article{Subramanian:1997gi,
    author = "Subramanian, Kandaswamy and Barrow, John D.",
    title = "{Magnetohydrodynamics in the early universe and the damping of noninear Alfven waves}",
    eprint = "astro-ph/9712083",
    archivePrefix = "arXiv",
    doi = "10.1103/PhysRevD.58.083502",
    journal = "Phys. Rev. D",
    volume = "58",
    pages = "083502",
    year = "1998"
}

@article{Sigl:1996dm,
    author = "Sigl, Guenter and Olinto, Angela V. and Jedamzik, Karsten",
    title = "{Primordial magnetic fields from cosmological first order phase transitions}",
    eprint = "astro-ph/9610201",
    archivePrefix = "arXiv",
    doi = "10.1103/PhysRevD.55.4582",
    journal = "Phys. Rev. D",
    volume = "55",
    pages = "4582--4590",
    year = "1997"
}

@article{Vachaspati:1991nm,
    author = "Vachaspati, T.",
    title = "{Magnetic fields from cosmological phase transitions}",
    doi = "10.1016/0370-2693(91)90051-Q",
    journal = "Phys. Lett. B",
    volume = "265",
    pages = "258--261",
    year = "1991"
}

@article{Enqvist:1993np,
    author = "Enqvist, K. and Olesen, P.",
    title = "{On primordial magnetic fields of electroweak origin}",
    eprint = "hep-ph/9308270",
    archivePrefix = "arXiv",
    reportNumber = "NBI-HE-93-33",
    doi = "10.1016/0370-2693(93)90799-N",
    journal = "Phys. Lett. B",
    volume = "319",
    pages = "178--185",
    year = "1993"
}

@article{Ellis:2019tjf,
    author = "Ellis, John and Fairbairn, Malcolm and Lewicki, Marek and Vaskonen, Ville and Wickens, Alastair",
    title = "{Intergalactic Magnetic Fields from First-Order Phase Transitions}",
    eprint = "1907.04315",
    archivePrefix = "arXiv",
    primaryClass = "astro-ph.CO",
    reportNumber = "KCL-PH-TH/2019-60, CERN-TH-2019-104",
    doi = "10.1088/1475-7516/2019/09/019",
    journal = "JCAP",
    volume = "09",
    pages = "019",
    year = "2019"
}

@article{Balaji:2024rvo,
    author = "Balaji, Shyam and Fairbairn, Malcolm and Olea-Romacho, Maria Olalla",
    title = "{Magnetogenesis with gravitational waves and primordial black hole dark matter}",
    eprint = "2402.05179",
    archivePrefix = "arXiv",
    primaryClass = "hep-ph",
    doi = "10.1103/PhysRevD.109.075048",
    journal = "Phys. Rev. D",
    volume = "109",
    number = "7",
    pages = "075048",
    year = "2024"
}

@article{Olea-Romacho:2023rhh,
    author = "Olea-Romacho, Mar{\'\i}a Olalla",
    title = "{Primordial magnetogenesis in the two-Higgs-doublet model}",
    eprint = "2310.19948",
    archivePrefix = "arXiv",
    primaryClass = "hep-ph",
    doi = "10.1103/PhysRevD.109.015023",
    journal = "Phys. Rev. D",
    volume = "109",
    number = "1",
    pages = "015023",
    year = "2024"
}

@article{Turner:1987bw,
    author = "Turner, Michael S. and Widrow, Lawrence M.",
    title = "{Inflation Produced, Large Scale Magnetic Fields}",
    reportNumber = "FERMILAB-PUB-87-154-A",
    doi = "10.1103/PhysRevD.37.2743",
    journal = "Phys. Rev. D",
    volume = "37",
    pages = "2743",
    year = "1988"
}

@article{Ratra:1991bn,
    author = "Ratra, Bharat",
    title = "{Cosmological 'seed' magnetic field from inflation}",
    reportNumber = "CALT-68-1750, GRP-286",
    doi = "10.1086/186384",
    journal = "Astrophys. J. Lett.",
    volume = "391",
    pages = "L1--L4",
    year = "1992"
}

@ARTICLE{2025A&A...698A.302L,
       author = {{Llerena}, M. and {Pentericci}, L. and {Napolitano}, L. and {Mascia}, S. and {Amor{\'\i}n}, R. and {Calabr{\`o}}, A. and {Castellano}, M. and {Cleri}, N.~J. and {Giavalisco}, M. and {Grogin}, N.~A. and {Hathi}, N.~P. and {Hirschmann}, M. and {Koekemoer}, A.~M. and {Nanayakkara}, T. and {Pacucci}, F. and {Shen}, L. and {Wilkins}, S.~M. and {Yoon}, I. and {Yung}, L.~Y.~A. and {Bhatawdekar}, R. and {Lucas}, R.~A. and {Wang}, X. and {Arrabal Haro}, P. and {Bagley}, M.~B. and {Finkelstein}, S.~L. and {Kartaltepe}, J.~S. and {Merlin}, E. and {Papovich}, C. and {Pirzkal}, N. and {Santini}, P.},
        title = "{The ionizing photon production efficiency of star-forming galaxies at z {\ensuremath{\sim}} 4{\textendash}10}",
      journal = {\aap},
         year = 2025,
        month = jun,
       volume = {698},
          eid = {A302},
        pages = {A302},
          doi = {10.1051/0004-6361/202453251},
archivePrefix = {arXiv},
       eprint = {2412.01358},
 primaryClass = {astro-ph.GA},
       adsurl = {https://ui.adsabs.harvard.edu/abs/2025A&A...698A.302L}
}

@ARTICLE{1999ApJ...514..648M,
       author = {{Madau}, Piero and {Haardt}, Francesco and {Rees}, Martin J.},
        title = "{Radiative Transfer in a Clumpy Universe. III. The Nature of Cosmological Ionizing Sources}",
      journal = {\apj},
         year = 1999,
        month = apr,
       volume = {514},
       number = {2},
        pages = {648-659},
          doi = {10.1086/306975},
archivePrefix = {arXiv},
       eprint = {astro-ph/9809058},
 primaryClass = {astro-ph},
       adsurl = {https://ui.adsabs.harvard.edu/abs/1999ApJ...514..648M}
}

@ARTICLE{2012ApJ...747..100S,
       author = {{Shull}, J. Michael and {Harness}, Anthony and {Trenti}, Michele and {Smith}, Britton D.},
        title = "{Critical Star Formation Rates for Reionization: Full Reionization Occurs at Redshift z {\ensuremath{\approx}} 7}",
      journal = {\apj},
         year = 2012,
        month = mar,
       volume = {747},
       number = {2},
          eid = {100},
        pages = {100},
          doi = {10.1088/0004-637X/747/2/100},
       adsurl = {https://ui.adsabs.harvard.edu/abs/2012ApJ...747..100S}
}

@ARTICLE{2013ApJ...768...71R,
       author = {{Robertson}, Brant E. and {Furlanetto}, Steven R. and {Schneider}, Evan and {Charlot}, Stephane and {Ellis}, Richard S. and {Stark}, Daniel P. and {McLure}, Ross J. and {Dunlop}, James S. and {Koekemoer}, Anton and {Schenker}, Matthew A. and {Ouchi}, Masami and {Ono}, Yoshiaki and {Curtis-Lake}, Emma and {Rogers}, Alexander B. and {Bowler}, Rebecca A.~A. and {Cirasuolo}, Michele},
        title = "{New Constraints on Cosmic Reionization from the 2012 Hubble Ultra Deep Field Campaign}",
      journal = {\apj},
         year = 2013,
        month = may,
       volume = {768},
       number = {1},
          eid = {71},
        pages = {71},
          doi = {10.1088/0004-637X/768/1/71},
archivePrefix = {arXiv},
       eprint = {1301.1228},
 primaryClass = {astro-ph.CO},
       adsurl = {https://ui.adsabs.harvard.edu/abs/2013ApJ...768...71R}
}

@ARTICLE{2025arXiv250701096G,
       author = {{Giovinazzo}, Emma and {Oesch}, Pascal A. and {Weibel}, Andrea and {Meyer}, Romain A. and {Witten}, Callum and {Bhagwat}, Aniket and {Brammer}, Gabriel and {Chisholm}, John and {de Graaff}, Anna and {Gottumukkala}, Rashmi and {Jecmen}, Michelle and {Katz}, Harley and {Leja}, Joel and {Marques-Chaves}, Rui and {Maseda}, Michael and {Shivaei}, Irene and {Trebitsch}, Maxime and {Verhamme}, Anne},
        title = "{Breaking Through the Cosmic Fog: JWST/NIRSpec Constraints on Ionizing Photon Escape in Reionization-Era Galaxies}",
      journal = {arXiv e-prints},
         year = 2025,
        month = jul,
          eid = {arXiv:2507.01096},
        pages = {arXiv:2507.01096},
          doi = {10.48550/arXiv.2507.01096},
archivePrefix = {arXiv},
       eprint = {2507.01096},
 primaryClass = {astro-ph.GA},
       adsurl = {https://ui.adsabs.harvard.edu/abs/2025arXiv250701096G}
}

@article{Ellis:2025xju,
    author = "Urrutia, Juan and Ellis, John and Fairbairn, Malcolm and Vaskonen, Ville",
    title = "{Starlight from JWST: Implications for star formation and dark matter models}",
    eprint = "2504.20043",
    archivePrefix = "arXiv",
    primaryClass = "astro-ph.CO",
    reportNumber = "KCL-PH-TH/2025-12, CERN-TH-2025-085, AION-REPORT/2025-03",
    doi = "10.1051/0004-6361/202555390",
    journal = "Astron. Astrophys.",
    volume = "702",
    pages = "A109",
    year = "2025"
}

@article{Paoletti:2022gsn,
    author = "Paoletti, D. and Chluba, J. and Finelli, F. and Rubi{\~n}o-Martin, J. A.",
    title = "{Constraints on primordial magnetic fields from their impact on the ionization history with Planck 2018}",
    eprint = "2204.06302",
    archivePrefix = "arXiv",
    primaryClass = "astro-ph.CO",
    doi = "10.1093/mnras/stac2947",
    journal = "Mon. Not. Roy. Astron. Soc.",
    volume = "517",
    number = "3",
    pages = "3916--3927",
    year = "2022"
}

\end{document}